\documentclass[journal=jacsat,manuscript=article]{achemso}

\usepackage{chemformula} 
\usepackage[T1]{fontenc} 

\author{Hui-Fei Zhai}
\affiliation{School of Molecular Sciences, Arizona State University, Tempe, Arizona 85287, United States}
\author{Sergey L. Bud'ko}
\affiliation{Ames National Laboratory, Iowa State University, Ames, Iowa 50011, United States}
\alsoaffiliation{Department of Physics and Astronomy, Iowa State University, Ames, Iowa 50011, United States}
\author{Jacob W. Fritsky}
\affiliation{School of Molecular Sciences, Arizona State University, Tempe, Arizona 85287, United States}
\author{Jason F. Khoury}
\affiliation{School of Molecular Sciences, Arizona State University, Tempe, Arizona 85287, United States}

\email{jason.khoury@asu.edu}

\title
  {Mixed-Valent Ce Disrupts Magnetic Ordering in CeFe$_2$Ga$_8$}

\begin{document}


\begin{abstract}
Mixed valency in intermetallics with lanthanide cations is well established as a pathway to unusual charge transport, complex magnetism, and superconductivity. In this work, we report a comprehensive study of the structural, magnetic, electronic, and thermal properties of the mixed valent compound CeFe$_2$Ga$_8$. Powder X-ray diffraction (PXRD) and X-ray photoelectron spectroscopy (XPS) characterize CeFe$_2$Ga$_8$ as a quasi-one-dimensional (Q1D) compound with mixed-valent Ce$^{3+}$ and Ce$^{4+}$ on a single crystallographic site. $^{57}$Fe M\"ossbauer spectroscopy indicates that the Fe sublattice is nonmagnetic, in direct contrast with recent reports of this compound. Low-temperature electrical resistivity and heat capacity measurements show no evidence of magnetic ordering, and a modest Sommerfeld coefficient ($\gamma$) of 22.7 mJ/mol$\cdot$K$^2$ make extensive Kondo hybridization unlikely. DC and AC magnetic susceptibility data suggest short-range magnetic order at $\sim$5.2 and 7.6 K with no frequency dependence, ruling out canonical spin-glass behavior in this compound. Additionally, the magnetic susceptibility data does not contain any broad features that are typically associated with an intermediate valence state in Ce, suggesting either high-temperature valence fluctuation or a different mechanism of mixed valency. This work demonstrates that mixed-valent Ce inhibits magnetic ordering in CeFe$_2$Ga$_8$ and provides a broader picture for how to analyze short-range spin interactions in Q1D intermetallics.
\end{abstract}

\section{Introduction}

Mixed-valent materials have long been of interest to solid-state chemists due to their tunable physical properties and versatile chemical bonding environments.\cite{woodward2003mixed} Typically, mixed-valent solid-state materials are divided into three categories. The first category represents compounds with distinct coordination environments for each oxidation state (I). The second category features compounds with similar coordination environments, but the mixed valency occurs due to electron hopping between them (II). Lastly, the third category involves solid-state structures with sufficiently delocalized electrons, causing all coordination environments to conform to an ``intermediate'' fractional valence state (III). In intermetallic materials, mixed valency can be a powerful knob for tuning physical properties, such as isolated magnetic sublattices, superconductivity, and heavy fermion behavior.\cite{wakiya2017structural, wakiya2019intermediate, gignoux1983intermediate, swatek2013intermediate, chondroudi2007mixed, bauminger1978mixed} However, despite the great promise of these different quantum material properties, the chemical bonding explanation for why mixed valency occurs in some materials but not others is still poorly understood. Developing a greater chemical understanding of mixed valency in intermetallics will ameliorate this issue, allowing for more intentional targeting of specific physical properties.

To further explore mixed valency in intermetallic systems, we must identify a versatile family of compounds where certain elements can be assigned oxidation states. The LnM$_2$X$_8$ structure (where Ln = lanthanide, M = transition metal, X = main group element) is one such example of a versatile family to investigate how lanthanide oxidation states affect magnetic behavior. These materials typically crystallize in the orthorhombic CaCo$_2$Al$_8$-type structure with space group \textit{Pbam} (No. 55), where the Ln ions form quasi-one-dimensional (Q1D) chains along the $c$-axis~\cite{sichevich1985,Ogunbunmi2021}. The Ln$^{3+}$ cation only has one crystallographic site, allowing for an investigation of Type II and III mixed valency. The LnM$_2$X$_8$ structure exhibits myriad magnetic and electronic behaviors that depend critically on the lanthanide oxidation state, with examples ranging from complex magnetism, superconductivity, and correlated electron behavior.~\cite{Wang2017npj, Ghosh2012, Fritsch2004, Ogunbunmi2018, Nair2017, Ogunbunmi2021, Cheng2019, Bhattacharyya2020prb, Zheng2022, Cheng2022, Schroder2000, Ogunbunmi2020, Wang2021, Wang2019}. 

For Fe-based compounds within this family, prior studies have often indicated that the Fe sublattice is not magnetic. For example, M\"{o}ssbauer spectroscopy measurements on CeFe$_2$Al$_8$ and LaFe$_2$Al$_8$ have demonstrated that Fe atoms are non-magnetic across wide temperature ranges, with the former compound exhibiting mixed-valent Ce~\cite{Tamura2000}. Neutron diffraction investigations on PrFe$_2$Al$_8$ confirmed long-range magnetic order of Pr below 4.5~K with the magnetic structure attributed to the Pr sublattice~\cite{Nair2017}. In the LnFe$_2$Ga$_8$ series, PrFe$_2$Ga$_8$ orders antiferromagnetically originating from Pr$^{3+}$ moments below 14~K~\cite{Wang2022a}, while some reports discuss the possibility of additional magnetic features~\cite{Ogunbunmi2020}. NdFe$_2$Ga$_8$ presents a complex magnetic landscape with two magnetic transitions at 10~K and 14.5~K~\cite{Wang2021,Wang2022b}, where neutron diffraction studies confirmed a commensurate AFM structure associated with Nd moments below 10~K, with incommensurate magnetic peaks observed between 10~K and 14.5~K~\cite{Wang2022b}. 

Recent work by Deng \textit{et al.} (2024)~\cite{Deng2024} reported on the physical properties of CeFe$_2$Ga$_8$ and concluded that Ce ions exhibit an intermediate valence state while also remaining non-magnetic at low temperatures. In their recent article, they attributed the observed magnetic behavior to short-range interactions involving the Fe sublattice rather than Ce moments, which contrasts with the non-magnetic Fe behavior observed in related compounds such as CeFe$_2$Al$_8$ and LaFe$_2$Al$_8$~\cite{Tamura2000}. Due to the precedent set by CeFe$_2$Al$_8$, we argue that the electronic and magnetic ground state of CeFe$_2$Ga$_8$ requires further investigation. Understanding whether magnetism originates from f-electrons (lanthanides) or d-electrons (transition metals) may provide some clues about the accessible parameter space for tuning between different quantum phases.

To resolve this apparent inconsistency with related Fe-based compounds in this family, we performed a comprehensive experimental investigation of CeFe$_2$Ga$_8$ using magnetic, thermal, electrical, and spectroscopic measurements. Herein, from M\"{o}ssbauer spectroscopy measurements, we show that the Fe sublattice remains non-magnetic throughout the measured temperature range, while X-ray Photoelectron Spectroscopy (XPS) suggests mixed-valent Ce between Ce$^{3+}$ and Ce$^{4+}$. DC Magnetic susceptibility measurements show ZFC-FC divergence below ~8 K, with features at $\sim$5.2 and 7.6 K that suggest short-range magnetic order. The fitted effective magnetic moments were 2.23 $\mu_B$ for $\mu_{0}H$ $\parallel$ \emph{c} and 2.04 $\mu_B$ for $\mu_{0}H$ $\perp$ \emph{c}, which are both smaller than the Ce$^{3+}$ theoretical value of 2.54 $\mu_B$. Surprisingly, there is no broad low-temperature feature in the magnetic susceptibility typical of an intermediate valence state, suggesting that the transition is either above room temperature or the mixed valency occurs due to a different mechanism. AC magnetic susceptibility measurements show no frequency dependence from the magnetic transitions, ruling out the possibility of spin glass behavior. Additionally, heat capacity measurements show no continuous phase transition typical of antiferromagnetic ordering. These measurements indicate that magnetic interactions in CeFe$_2$Ga$_8$ are primarily short-range Ce-based interactions as opposed to long range magnetic order, with no contribution from the Fe sublattice. This work demonstrates how mixed valency directly disrupts the magnetic behavior of Q1D intermetallics, as well as discussing the structural features that lead to this phenomenon.

\section{Experimental Methods}

\subsection{Synthesis}
High-quality single crystals of CeFe$_2$Ga$_8$ were synthesized using the Ga flux method. Starting materials of Ce (99.9\%, Thermo Scientific), Fe (99.99\% Sigma Aldrich), and Ga (99.999\%, Thermo Scientific) were mixed in a molar ratio of 1:2:20. Specifically, Ce (2 mmol, 0.2802 g), Fe (4 mmol, 0.2234 g) and excess Ga (40 mmol, 2.7889 g) were loaded into an alumina crucible, and then sealed in an evacuated quartz tube. Ce was handled in an Ar glove box to prevent oxidation. The reagents were heated to 1000°C, maintained for 10 hours, and then slowly cooled to 700°C at a rate of 2°C/h, followed by centrifugation to separate the crystals from the flux.

\subsection{X-ray Diffraction}
The crystal structure was confirmed by powder X-ray diffraction (XRD) of crushed single crystals performed on a Bruker D6 Phaser with CuK$\alpha$ radiation ($\lambda$ = 1.5406$\mathring{A}$) and a Ni K$\beta$ filter at room temperature, employing the voltage of 40 kV and the current of 15 mA. Data were collected from 10$^\circ$ to 90$^\circ$ (2$\theta$) with a step size of 0.05$^\circ$ and a dwell time of 3~seconds per step. The structure was determined by Rietveld refinement using the commercial TOPAS software. 

\subsection{X-ray Photoelectron Spectroscopy (XPS)}
X-ray photoelectron spectroscopy data was collected using the ground powder of CeFe$_2$Ga$_8$ single crystals on a Kratos Axis Supra+ with an Al K$\alpha$ source at room temperature. Wide survey scans were collected with a pass energy of 80 eV, while high-resolution scans of the Ce 3d region were acquired with a pass energy of 20 eV.

\subsection{Magnetometry}
The magnetic susceptibility ($\chi$) of CeFe$_2$Ga$_8$ (3.0 mg) was measured applying a field of 25, 100 and 1000 Oe both along and perpendicular to the $c$ direction under zero field cooling (ZFC) and field cooling (FC) mode, using a Quantum Design Physical Property Measurement System (PPMS) equipped with a vibrating sample magnetometer (VSM). Magnetic isotherms of CeFe$_2$Ga$_8$ was measured from -9 T to 9 T at various temperatures from 2 K to 300 K, along two crystallographic directions. AC susceptibility measurements were conducted with using the AC option of the Quantum Design Magnetic Property Measurement System (MPMS3) an AC field amplitude of 3 Oe at frequencies from 3.25 to 925.7 Hz to probe the dynamic magnetic behavior.

\subsection{Resistivity}
The electrical resistivity was measured on CeFe$_2$Ga$_8$ single crystals using Quantum Design PPMS from 2 to 300 K with dimensions of 0.17 $\times$ 0.20 $\times$ 0.62 mm. Four Au wires were attached linearly to the surface of the crystals using DuPont silver paint with current along the $c$ direction while the magnetic field (from -9 T to 9 T) was applied  perpendicular to the $c$ direction.

\subsection{M\"ossbauer Spectroscopy}

$^{57}$Fe M\"ossbauer spectroscopy measurements were performed using a SEE Co. conventional constant acceleration type spectrometer in transmission geometry with an $^{57}$Co (Rh) source, kept at room temperature. The absorber was cooled to a desired temperature using a Janis model SHI-850-5 closed cycle refrigerator (with vibration damping). The driver velocity was calibrated by $\alpha$-Fe foil and all isomer shifts (IS) are quoted relative to the $\alpha$-Fe foil at room temperature. A commercial software package MossWinn was used to analyze the M\"ossbauer spectra.

\subsection{Heat Capacity}
The heat capacity was measured using a Quantum Design PPMS between 1.8 and 300 K with the thermal-relaxation method. A single crystal with a mass of 2.6 mg was attached to the sample stage using Apiezon N grease. To investigate potential field-dependent magnetic contributions, additional low-temperature measurements (2–10 K) were performed under applied magnetic fields of 0, 1, and 3 T perpendicular to the $c$-axis.

\section{Results and Discussion}

\subsection{Crystal Structure}
\begin{figure*}
   \centering
    \includegraphics[width=\linewidth]{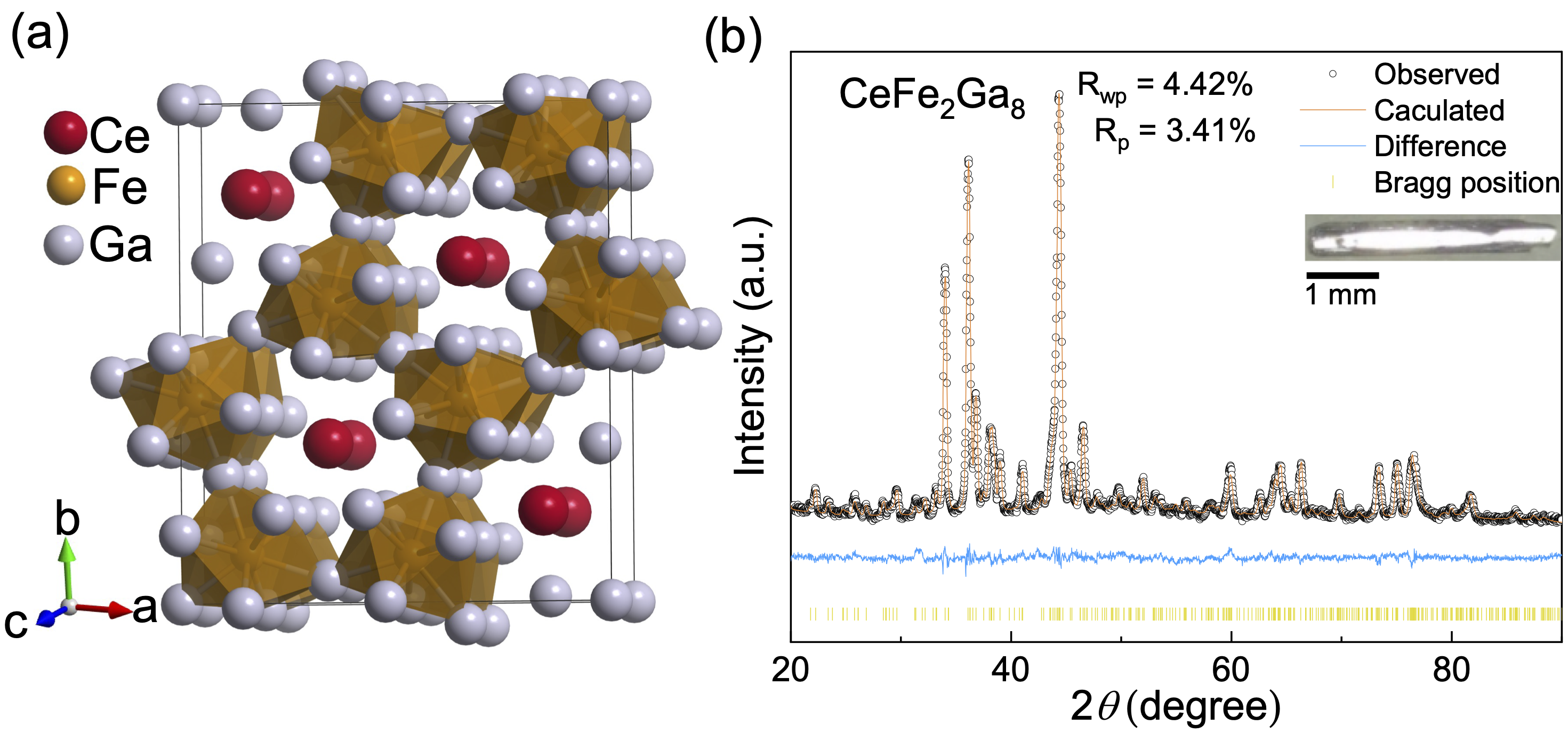}
    \caption{The unit cell of CeFe$_2$Ga$_8$ (a) which crystallizes in orthorhombic symmetry in the \emph{Pbam} space group. Rietveld refinement of the powder X-ray diffraction (PXRD) pattern (b) confirms the CeFe$_2$Al$_8$ structure type. The inset shows a typical single crystal with a 1 mm scale bar.}
    \label{fig:structure}
\end{figure*}

Powder X-ray diffraction (PXRD) of crushed single crystals confirmed that CeFe$_2$Ga$_8$ crystallizes in orthorhombic symmetry in the \emph{Pbam} space group (No. 55) and the CeFe$_2$Al$_8$ structure type, with lattice parameters $a = 12.4718(8)$ Å, $b = 14.3711(6)$ Å, $c = 4.0743(7)$ Å, consistent with previous reports~\cite{Deng2024}. As shown in Fig.~\ref{fig:structure}, CeFe$_2$Ga$_8$ features Ce atoms arranged in Q1D chains along the c-axis, surrounded by corner and face-sharing FeGa$_9$ polyhedra. The Fe--Ga bond lengths in the polyhedra range from 2.3546(5) -- 2.6603(5) Å, comparable to the same bonds in binary FeGa$_3$ (2.3646(4) -- 2.4996(8) Å), which consists of FeGa$_8$ polyhedra that also exhibit corner and face-sharing connectivity.\cite{haussermann2002fega3} However, Fe--Ga bond lengths tend to be longer in other binaries such as Fe$_4$Ga$_3$ (2.602(4) -- 2.651(4) Å), which is comprised of only face-sharing FeGa$_8$ polyhedra.\cite{philippe1975structures} Since the Fe--Ga bond lengths in Fe$_4$Ga$_3$ tend to be more similar to the longer bonds in CeFe$_2$Ga$_8$, it is possible that its higher coordination number and variable polyhedral connectivity contribute to the several different bonds that are found in the structure.

The intra-chain Ce–-Ce distance is 4.07 Å, while the inter-chain separations are significantly larger at 6.54 Å and 7.55 Å, establishing the one-dimensional character of the magnetic lanthanide sublattice. 
The quasi-1D crystalline anisotropy creates well-separated Ce chains that may promote local spin frustration. Fig.~\ref{fig:structure}b shows the Rietveld refinement of the X-ray diffraction pattern, with strong agreement between the observed and calculated data ($R_{\text{wp}} = 4.42 \%$, $R_\text{p} = 3.41 \%$). The Ce atoms in the structure are on a single crystallographic Wyckoff position, which simplifies the picture of trying to understand the magnetic contributions of the lanthanide sublattice in CeFe$_2$Ga$_8$. An investigation into whether or not Ce and Fe are moment bearing are needed to definitively explain the behavior of this compound, which we expand upon in the spectroscopy and magnetometry sections below.

\subsection{X-ray Photoelectron Spectroscopy}
\begin{figure}
    \centering
    \includegraphics[width=12cm]{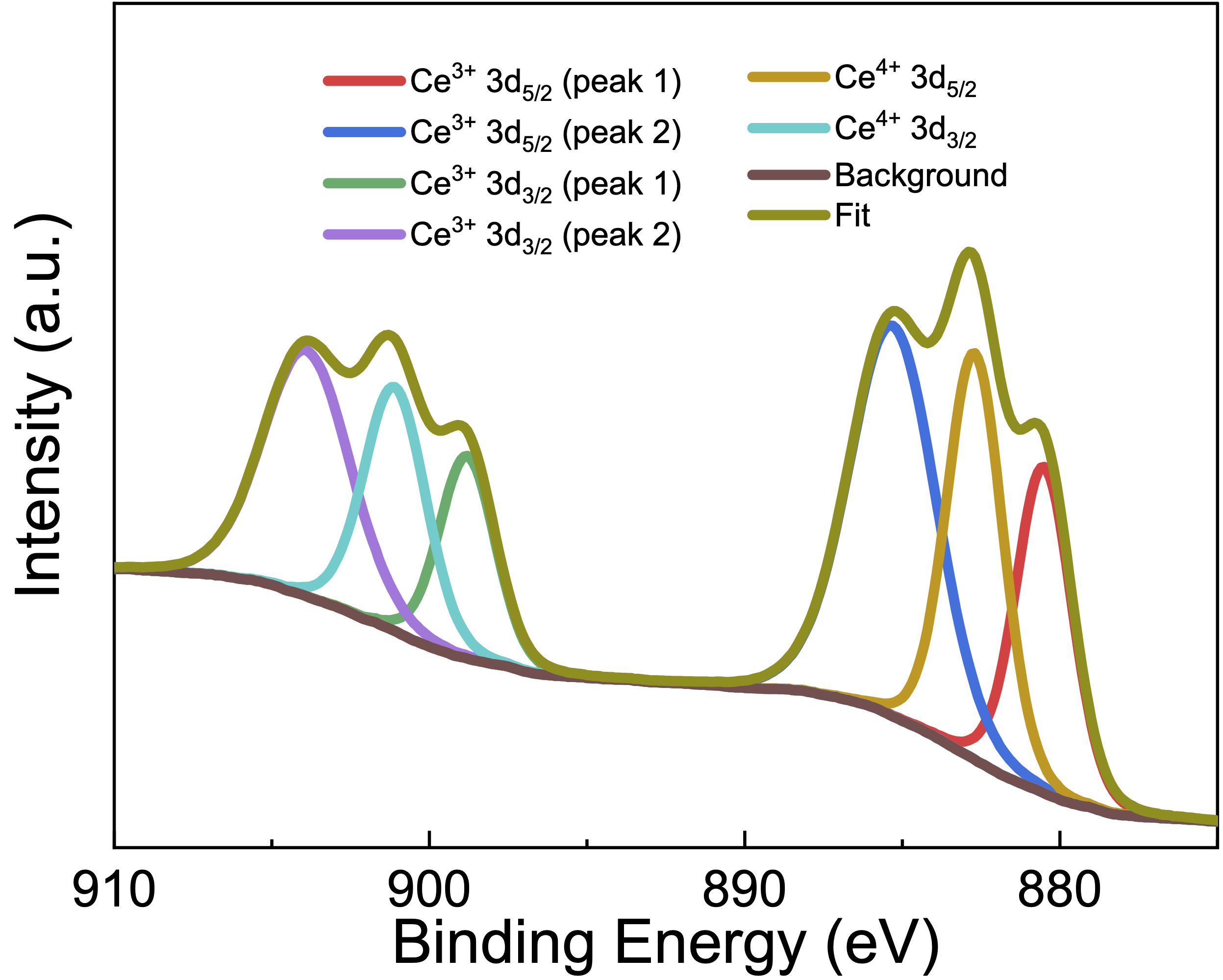}
    \caption{X-ray photoelectron spectroscopy (XPS) of Ce 3$d$ core levels in CeFe$_2$Ga$_8$. The fitted spectrum shows Ce$^{3+}$ components including 3d$_{5/2}$ peaks and 3d$_{3/2}$  peaks, along with Ce$^{4+}$ 3d$_{5/2}$ and 3d$_{3/2}$ peaks. The data indicate mixed-valent Ce in a 7:3 ratio between Ce$^{3+}$ and Ce$^{4+}$.}
    \label{fig:XPS}
\end{figure}

Given the ambiguity surrounding the origin of magnetic correlations in CeFe$_2$Ga$_8$, we performed X-ray photoelectron spectroscopy (XPS) measurements to gain insight into the coordination environment of Ce in the structure. Figure~\ref{fig:XPS} shows the Ce XPS spectra of CeFe$_2$Ga$_8$, consisting of 3$d$$_{3/2}$ and 3$d$$_{5/2}$ peaks in the 3$d$ core level. The spectra have to be fitted with two distinct Ce atoms to account for the peak shoulders, and since there is only one unique crystallographic site for Ce in CeFe$_2$Ga$_8$, it is likely that there is mixed valency between Ce$^{3+}$ and Ce$^{4+}$ in the structure. The fitted binding energies are also in line with literature precedent of mixed valency between Ce$^{3+}$ and Ce$^{4+}$ in other solid-state materials.\cite{renaudin2010pseudotetragonal} For the atom assigned Ce$^{3+}$, the 3d$_{5/2}$ peaks have binding energies of $\sim$885 eV and 880 eV, while the 3d$_{3/2}$ peaks appear at $\sim$899 eV and 904 eV, respectively. The Ce$^{4+}$ atom is fitted by peaks at$\sim$882 eV (3d$_{5/2}$) and $\sim$901 eV (3d$_{3/2}$). The fitted results suggest a $\sim$7:3 ratio Ce$^{3+}$ and Ce$^{4+}$. The mixed-valent Ce on the sole Wyckoff position in the structure naturally creates a random distribution of magnetic (Ce$^{3+}$) and non-magnetic (Ce$^{4+}$) 4$f$ cations throughout the lattice. In conjunction with the magnetometry and M\"ossbauer spectroscopy data described in the sections below, these results support a mixed-valence state where only Ce$^{3+}$ (4f$^1$) contributes to the magnetic moment, leading to magnetic correlations but preventing long range ordering in the lattice.

\subsection{M\"ossbauer Spectroscopy}

\begin{figure*}
    \centering
    \includegraphics[width=\linewidth]{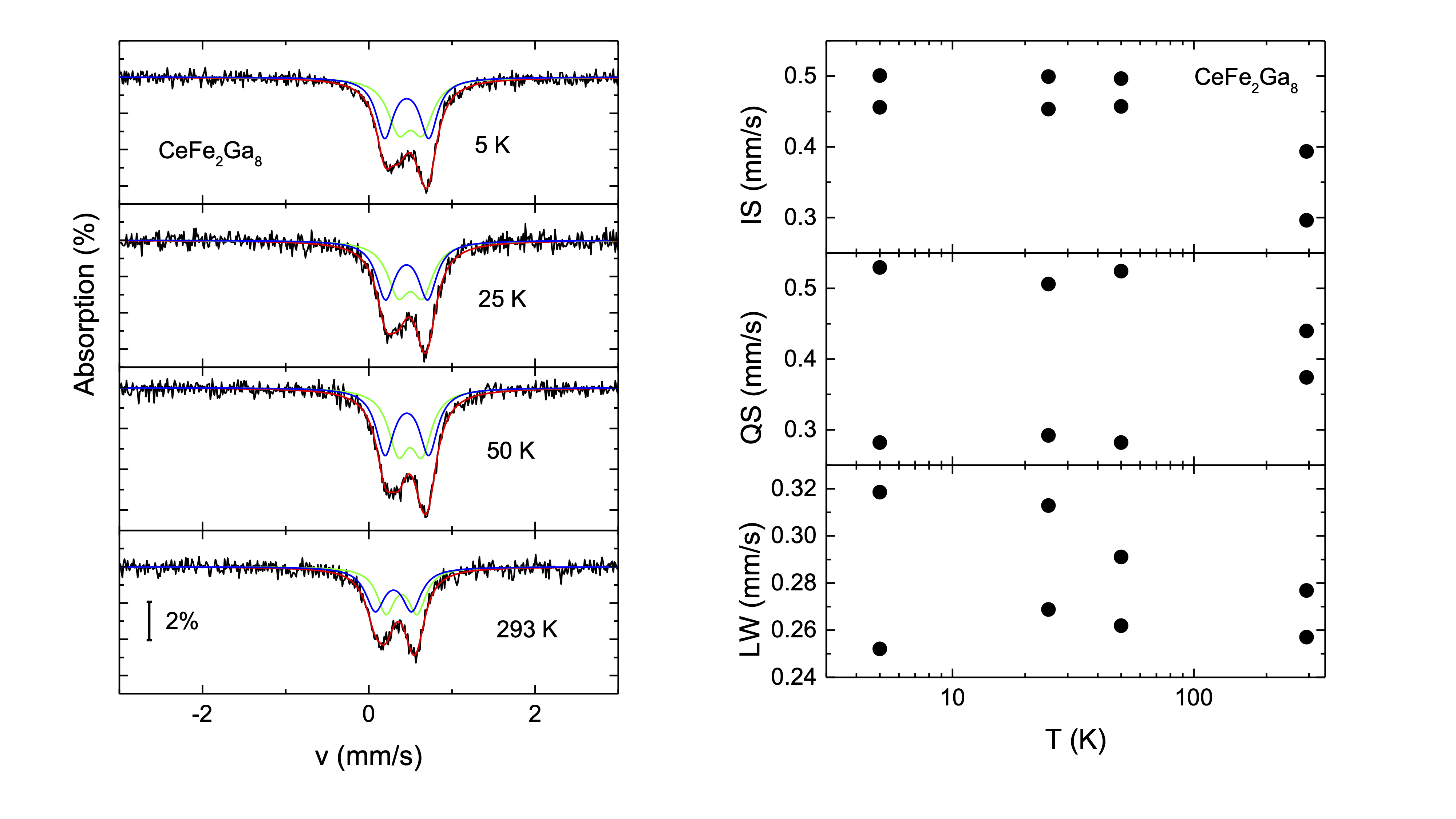}
    \caption{$^{57}$Fe Mössbauer spectroscopy data for CeFe$_2$Ga$_8$ showing absorption spectra at different temperatures (left) and temperature dependence of the hyperfine parameters (right): Isomer Shift(IS), Quadrupole Splitting(QS), Line Width(LW).}
    \label{fig:Mossbauer}
\end{figure*}

Given the possibility of mixed-valent Ce in CeFe$_2$Ga$_8$, we performed M\"ossbauer spectroscopy to determine whether or not the Fe sublattice was moment bearing. The $^{57}$Fe M\"ossbauer spectra in Fig.~\ref{fig:Mossbauer} demonstrate that Fe atoms remain in a non-magnetic state down to 5 K, as evidenced by the absence of magnetic hyperfine splitting. In addition, the transferred hyperfine field on $^{57}$Fe is small enough not to be detected in the background of the quadrupolar splitting. Such behavior is consistent with observations in related compounds including CeFe$_2$Al$_8$, LaFe$_2$Al$_8$, and YbNi$_{2-x}$Fe$_x$Al$_8$\cite{Tamura2000,Ogunbunmi2021}, further suggesting the absence of magnetic moments on the Fe sublattice in this family of materials. Similar to other Fe-based compounds in the LnM$_2$X$_8$ family, such as LaFe$_2$Al$_8$, CeFe$_2$Al$_8$\cite{Tamura2000}, PrFe$_2$Al$_8$\cite{Nair2017} and NdFe$_2$Ga$_8$\cite{Wang2022b}, our M\"ossbauer spectroscopy measurements confirm that the Fe atoms remain non-magnetic, while the magnetic interactions in CeFe$_2$Ga$_8$ originate from the Ce sublattice.

\subsection{Electrical Resistivity}
\begin{figure*}
    \centering
    \includegraphics[width=\linewidth]{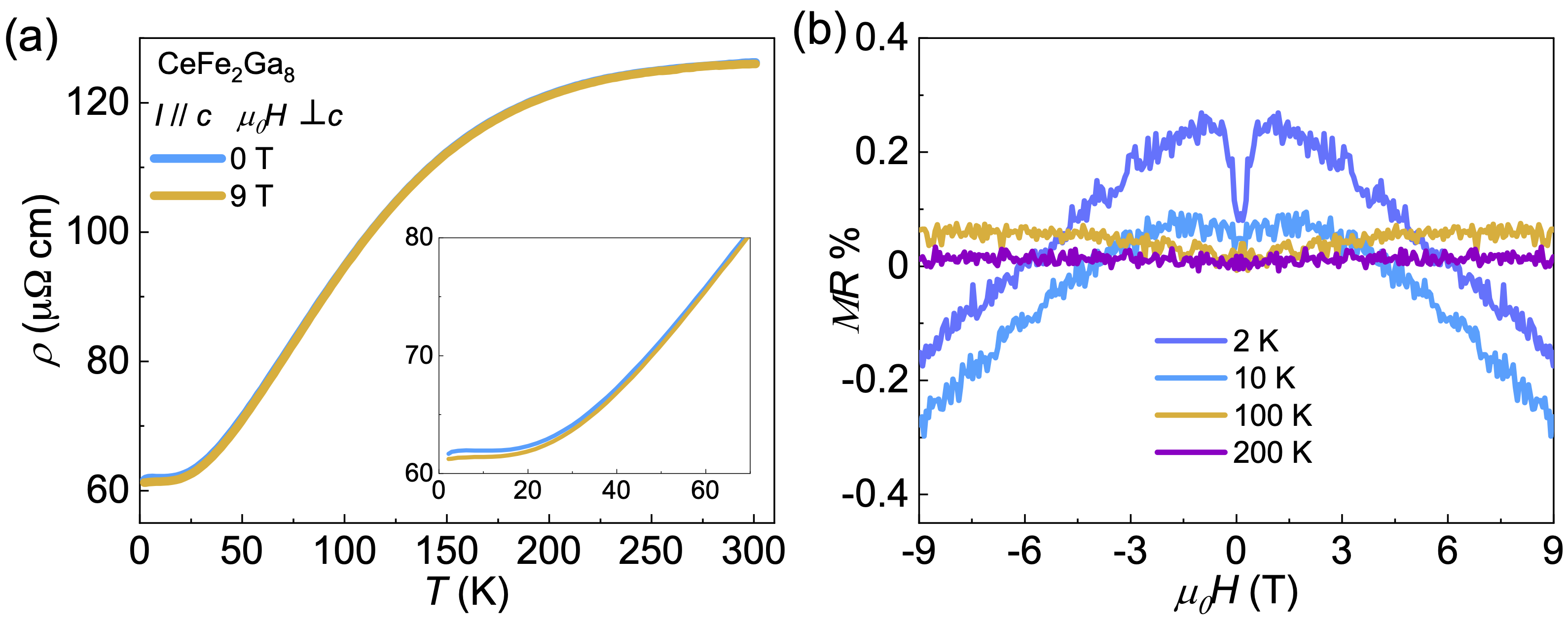}
    \caption{Electrical transport properties of CeFe$_2$Ga$_8$: (a) Resistivity ($\rho$) vs temperature with inset showing the low-temperature region. (b) Variable-temperature magnetoresistance.}
    \label{fig:resistivity}
\end{figure*}

Now that we have established mixed-valent Ce and nonmagnetic Fe in CeFe$_2$Ga$_8$, electrical resistivity measurements were performed on single crystals to observe evidence of low-temperature phenomena. Figure~\ref{fig:resistivity}(a) displays the electrical resistivity ($\rho$) versus temperature for CeFe$_2$Ga$_8$ with current and applied magnetic field parallel and perpendicular  to the $c$ direction, respectively. The resistivity exhibits metallic behavior over the entire temperature range, and gradually reaches a minimum value of $\sim$60 $\mu\Omega\ \text{cm}$ below approximately 10 K (inset of Figure~\ref{fig:resistivity}(a)). The charge transport behavior of CeFe$_2$Ga$_8$ is largely unchanged with an applied magnetic field of 9 T over the entire temperature range, indicative of minimal magnetoresistance from the compound. Notably, the charge transport data show no low-temperature events, except for broad, barely visible maximum at $\sim 5$~K, implying that there are no continuous phase transitions (such as long-range magnetic ordering or a structural phase transition) in CeFe$_2$Ga$_8$, which corroborates our heat capacity and magnetometry data in the sections below.

The field-dependent magnetoresistance (MR) at several temperatures is shown in Fig.~\ref{fig:resistivity}(b). At high temperatures( 100 and 200 K), the MR is essentially flat and featureless, as expected. As the temperature decreases, a weak, negative magnetoresistance emerges. At 2 K, we observe a small negative magnetoresistance of $\sim$--0.2\% at $\pm9$~T. The low magnitude of the MR indicates that the resistivity is robust to external magnetic fields, likely due to low charge-carrier mobility in the lattice. The weak negative value of the magnetoresistance is attributable to the local magnetic correlations in CeFe$_2$Ga$_8$, which cause spin-disorder scattering that is suppressed in the presence of an applied magnetic field. The relatively sharp dip at $\mu_{0}H = 0$ in 2~K magnetoresistance data is most likely related to the broad maximum in $\rho(T)$ mentioned above. Additional studies are required to understand the origin of these features.

\subsection{Heat Capacity}

\begin{figure*}
    \centering
    \includegraphics[width=\linewidth]{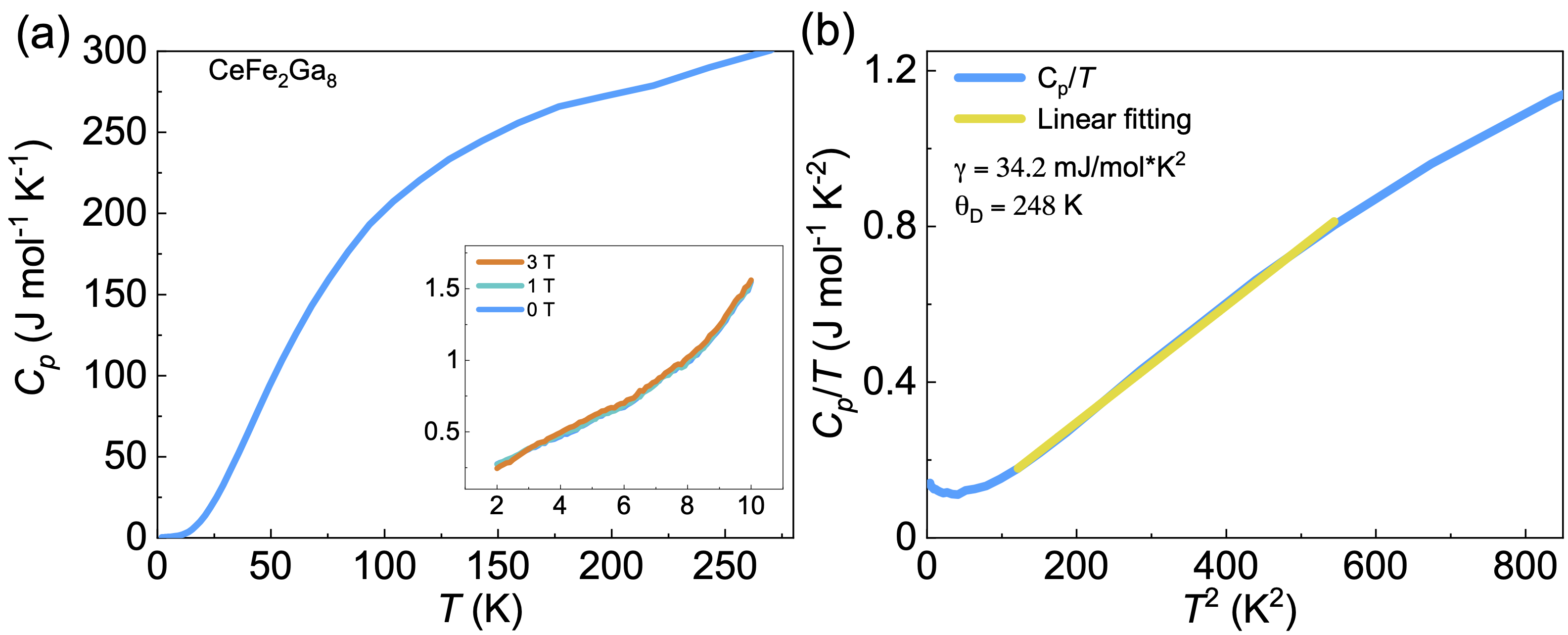}
    \caption{Heat capacity measurements of CeFe$_2$Ga$_8$. (a) $C_p$. The inset shows the low-temperature data with applied magnetic field at 0, 1 and 3 T. (b) $C_p/T$ vs $T^2$ analysis from 10 - 25 K yielding a Sommerfeld coefficient ($\gamma$) of 34.2 mJ/mol$\cdot$K$^2$ and a Debye temperature ($\theta_D$) of 248 K.}
    \label{fig:specificheat}
\end{figure*}

In concert with the electrical resistivity data, we performed heat capacity measurements to further support the lack of low-temperature continuous phase transitions in CeFe$_2$Ga$_8$, shown in Fig.~\ref{fig:specificheat}(a). The data show no lambda-like anomaly that would indicate long-range magnetic ordering at low temperatures, and this continues to be the case in variable-field data (shown in the inset of Fig.~\ref{fig:specificheat}(a)). The results altogether suggest short-range magnetic correlations from mixed-valent Ce, which is supported by the magnetometry data in the following section. The low-temperature specific heat was analyzed using a conventional model, in which the total specific heat is expressed as $C = \gamma T + \beta T^{3}$, where the first term represents the electronic contribution and the second term corresponds to the lattice contribution, as shown in Fig.~5(b). This analysis yields the electronic specific heat coefficient (Sommerfeld coefficient) $\gamma \approx 34.2 $ mJ/mol$\cdot$K$^2$, and $\beta \approx 1.5$ mJ/mol$\cdot$K$^4$ ($\theta_D = 248$ K). The fitted $\gamma$ value is significantly smaller than previously reported values for this compound~\cite{Deng2024}, suggesting a lower density of states (DOS) at the Fermi level.

\subsection{Magnetometry}
\begin{figure*}
    \centering
    \includegraphics[width=\linewidth]{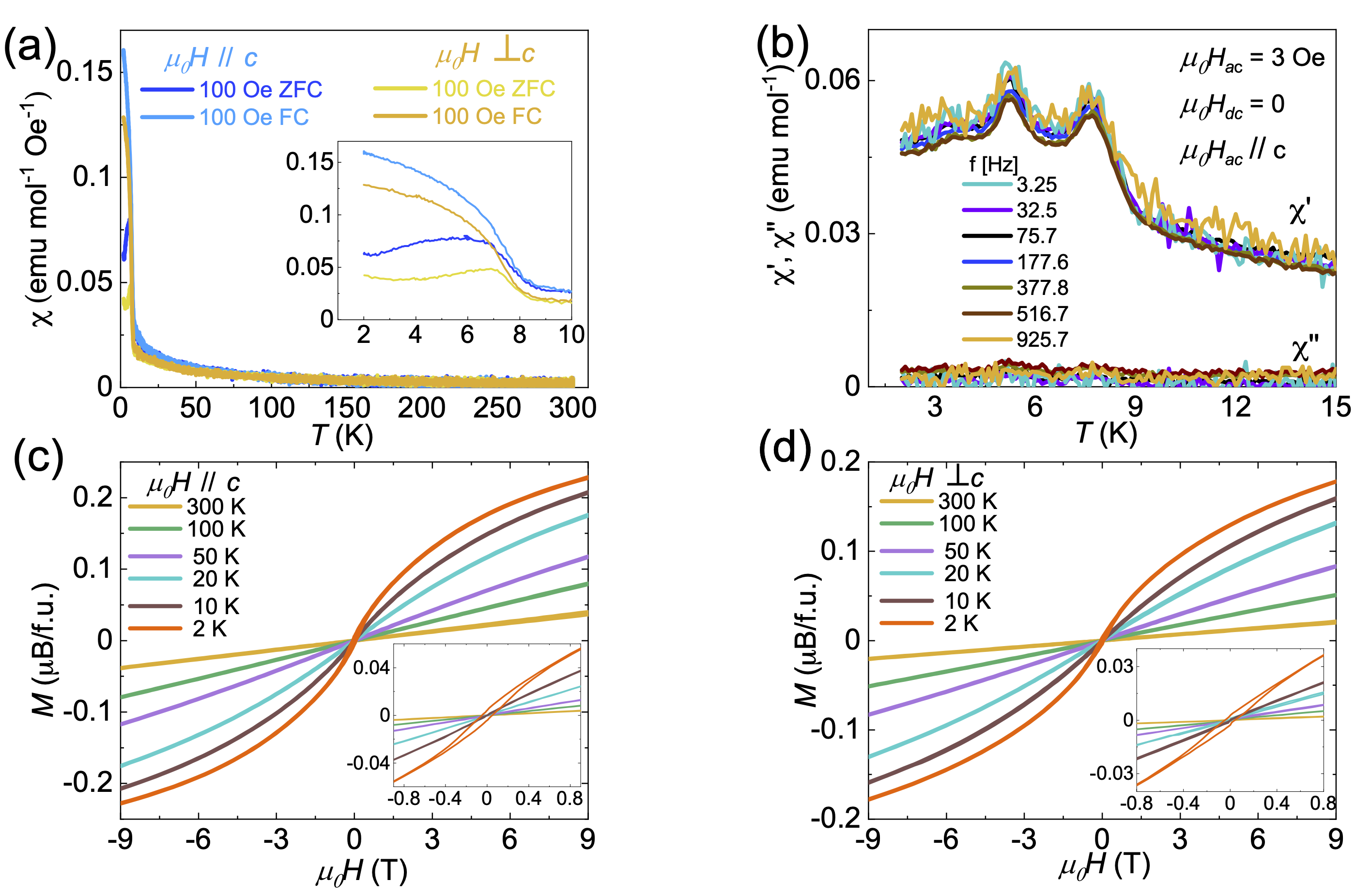}
    \caption{Magnetometry of CeFe$_2$Ga$_8$ : (a) DC magnetic susceptibility vs temperature for both zero-field-cooled (ZFC) and field-cooled (FC) data with $\mu_{0}H$ $\parallel$ and $\perp$ the \emph{c} direction.  The inset shows the low-temperature data. (b) AC susceptibility shows real part $\chi'$ and imaginary part $\chi''$ vs temperature at different frequencies with $\mu_{0}H$ $\parallel$ c. (c) and (d) magnetization curves at different temperatures with the inset showing the small hysteresis of the low field region, with $\mu_{0}H$ $\parallel$ and $\perp$ to the \emph{c} direction.}
    \label{fig:M}
\end{figure*}

Now that we have established the absence of any long-range magnetic ordering transition, the nonmagnetic character of the Fe sublattice, and the presence of mixed-valent Ce in CeFe$_2$Ga$_8$, we can be confident that our magnetometry measurements would only reflect magnetic correlations in Ce. If Ce in CeFe$_2$Ga$_8$ is mixed-valent  between Ce$^{3+}$ and Ce$^{4+}$, it should result in a lower magnetic moment and the suppression of long-range magnetic order from the lanthanide cations. To further investigate this possibility, we performed DC and AC magnetic susceptibility, shown in Fig.~\ref{fig:M}. The DC magnetic susceptibilty measurements show divergence between the zero-field-cooled (ZFC) and field-cooled (FC) data below $\sim$8 K with an applied field of 100 Oe, which is consistent when $\mu_{0}H$ is applied both $\parallel$ and $\perp$ to the \emph{c} direction. At higher temperatures (25-300 K), the magnetic susceptibility follows Curie-Weiss behavior for both orientations. The effective magnetic moments extracted from Curie-Weiss fits are 2.23 and 2.04 $\mu_B$ for $\mu_{0}H$ applied $\parallel$ and $\perp$ to the \emph{c} direction, respectively, which are both smaller than the theoretical value of 2.54 $\mu_B$ for Ce$^{3+}$. We can also estimate the polycrystalline-averaged susceptibility via the  relation $\chi_{\mathrm{avg}} = \frac{1}{3}(\chi_{\parallel} + 2\chi_{\perp})$, where the corresponding effective magnetic moment is approximately 2.10~$\mu_{\mathrm{B}}$.
The reduction in moment is consistent with the mixed valence between Ce$^{3+}$ and Ce$^{4+}$ suggested from our XPS measurements in the previous section. The negative Curie-Weiss temperatures ($\theta_C = $$\sim$$-24$ and $-31$ K for $\mu_{0}H$ $\parallel$ and $\perp$ to the \emph{c} direction, respectively) indicate predominantly antiferromagnetic interactions, and the comparable magnitude of both parameters suggests an absence of strong magnetic anisotropy in the system. 

The reduction in magnetic moment due to mixed-valent Ce in CeFe$_2$Ga$_8$ should stifle long-range magnetic ordering, since we have a combination of magnetic (Ce$^{3+}$) and nonmagnetic (Ce$^{4+}$) cations distributed randomly throughout the quasi-1D Ce chains. One possible consequence of this could be spin-glass behavior, where there is a random freezing of spins with no long-range magnetic ordering. To investigate this possibility, we performed AC magnetic susceptibility measurements (Fig.~\ref{fig:M}(b)) on a single crystal of CeFe$_2$Ga$_8$ with magnetic field applied along the \emph{c} direction. The real part ($\chi'$) exhibits two broad maxima at approximately 5.2 K and 7.6 K, respectively, while the imaginary part ($\chi''$) is nearly zero across all measured frequencies (3.25 -- 925.7 Hz) with minimal frequency dependence. These two peaks are consistent with our low-field DC magnetic susceptibility measurements at 25 Oe (shown in Fig.S2(a)), indicating highly field-sensitive magnetic correlations that were not observed in previous reports of CeFe$_2$Ga$_8$.\cite{Deng2024} As the applied field increases to 100 Oe and then to 1000 Oe (Fig.S2(b)),
these features become progressively suppressed and eventually merge into a single, broader transition. 

The AC magnetic susceptibility data do not show the characteristic frequency-dependent shift in magnetic peaks in $\chi'$ that are the typical ``smoking gun'' of classical spin glasses.\cite{Mugiraneza2022, Binder1986SpinGlassReview, Mydosh1993SpinGlass} In a spin glass, the response will depend on the frequency of the alternating field, where the freezing temperature shifts with frequency according to $T_f \propto (f/f_0)^{1/z\nu}$, with $f_0$ being the characteristic attempt frequency, and $z$ and $\nu$ are the dynamic and correlation length critical exponents, respectively.\cite{topping2018ac, Binder1986SpinGlassReview, Mydosh1993SpinGlass} Instead, the observed ZFC-FC divergence is consistent with short-range magnetic order rather than spin glass behavior, which is understandable given the assortment of magnetic (Ce$^{3+}$) and nonmagnetic (Ce$^{4+}$) cations. The nonmagnetic cations effectively act as magnetic blocking agents, preventing the long-range antiferromagnetic order typically seen with other lanthanides in this structure type.

The absence of clear spin-glass behavior, coupled with the presence of short-range magnetic order in this mixed-valent compound, suggests valence fluctuation typically seen in Ce-containing intermetallics that undergo Kondo hybridization.\cite{gignoux1983intermediate} However, there is no broad hump in the magnetic susceptibility data along either direction for CeFe$_2$Ga$_8$, which is common for intermediate valence state behavior in Ce-containing intermetallics.\cite{swatek2013intermediate, wakiya2017structural, wakiya2019intermediate, gignoux1983intermediate} One possibility is that the hump in susceptibility is present above room temperature, which is consistent with our room-temperature XPS data indicating mixed-valent Ce. Additionally, intermediate valence behavior tends to coincide with Kondo hybridization of Ce 4$f$ electrons, which conflicts with a low Sommerfeld coefficient from the fitted heat capacity data shown in the section above. It is possible that the mixed-valent Ce in CeFe$_2$Ga$_8$ is due to another mechanism, such as electron hopping typically seen in a Type-II mixed-valent system.\cite{woodward2003mixed} Another possibility could be local distortions around the Ce cations. Future studies of structural distortions from pair-distribution function (PDF) analysis would assist in attaining a definitive answer to this question.

Magnetization isotherms ($M$–$H$) for both field orientations (Fig.~\ref{fig:M}c and d) show minor hysteresis at low temperatures, suggesting weak ferromagnetic-like interactions. The presence of this small low-field hysteresis contrasts with the negative Weiss temperatures derived from Curie–Weiss fits, which indicate predominantly antiferromagnetic correlations. The apparent contradiction is explained by the mixed valency of the Ce sublattice: If there is no long range magnetic order, it is likely attributable to competing ferromagnetic and antiferromagnetic correlations between the Ce$^{3+}$ cations, with the latter being more prominent. In addition, the lack of magnetization saturation under applied magnetic fields up to 9 T supports the presence of short-range magnetic interactions rather than conventional long-range order. The lack of long-range magnetic order in CeFe$_2$Ga$_8$ stands in stark contrast to other isostructural compounds such as PrFe$_2$Al$_8$\cite{Nair2017} and NdFe$_2$Ga$_8$\cite{Wang2022b}, where their quasi-1D Ln$^{3+}$ chains exhibited long-range antiferromagnetic ordering. The combination of the absence of a sharp peak in the heat capacity, observed ZFC-FC splitting in the magnetic susceptibility, as well as minor hysteresis in magnetic moment vs. applied field data collectively provides evidence for a short-range magnetic ordering scenario, rather than long-range order or heavy fermion behavior.

\section{Conclusion}
In conclusion, we show a lack of long-range magnetic ordering in CeFe$_2$Ga$_8$ due to mixed-valent Ce, resulting in short-range magnetic interactions in the lanthanide sublattice. M\"ossbauer spectroscopy confirms the non-magnetic nature of Fe atoms, establishing Ce as the sole magnetic sublattice, contrary to previous reports~\cite{Deng2024}. XPS measurements provide direct evidence of mixed-valent Ce, with a ratio of $\sim$7:3 Ce$^{3+}$:Ce$^{4+}$, leading to a random distribution of magnetic and non-magnetic cations throughout the 1D Ce chain in the crystal structure. While the ZFC-FC splitting in the DC magnetic susceptibility initially suggested spin glass behavior, AC magnetic susceptibility measurements did not exhibit the frequency dependence expected for this phenomenon, further supporting short-range magnetic order rather than disordered spin freezing. The charge transport and heat capacity data corroborated this interpretation due to a lack of any low-temperature continuous phase transitions in CeFe$_2$Ga$_8$. These findings demonstrate how mixed valency in CeFe$_2$Ga$_8$ affects its magnetic behavior, highlighting the importance of comprehensive characterization using multiple complementary techniques. Future investigations using techniques such as neutron scattering or muon spin relaxation ($\mu$SR) would be valuable to further elucidate the nature of magnetic interactions in these Q1D intermetallic materials.

\begin{acknowledgement}
J.F.K. would like to acknowledge the School of Molecular Sciences at Arizona State University for providing start-up funding for this
work. In addition, we acknowledge the use of facilities within the Eyring Materials Center at Arizona State University. We acknowledge Dr. Yunbo Ou for assistance with the PPMS measurements and Dr. Pan Zhang for helpful discussions. Work done in Ames (SLB) was supported by the U.S. Department of Energy, Office of Basic Energy Science, Division of Materials Sciences and Engineering under Contract No. DE-AC02-07CH11358.

\end{acknowledgement}

\begin{suppinfo}

The Supporting Information is available:

\begin{itemize}
  \item Figure S1: Scanning Electron Microscopy (SEM) and Energy Dispersive Spectroscopy (EDS) of CeFe$_2$Ga$_8$
  \item Figure S2: ZFC-FC measurements of CeFe$_2$Ga$_8$ at 25 Oe and 1000 Oe
\end{itemize}

\end{suppinfo}


\providecommand{\latin}[1]{#1}
\makeatletter
\providecommand{\doi}
  {\begingroup\let\do\@makeother\dospecials
  \catcode`\{=1 \catcode`\}=2 \doi@aux}
\providecommand{\doi@aux}[1]{\endgroup\texttt{#1}}
\makeatother
\providecommand*\mcitethebibliography{\thebibliography}
\csname @ifundefined\endcsname{endmcitethebibliography}  {\let\endmcitethebibliography\endthebibliography}{}

For Table of Contents Only
\begin{figure}[H]
\centering
 \includegraphics[width=\textwidth]{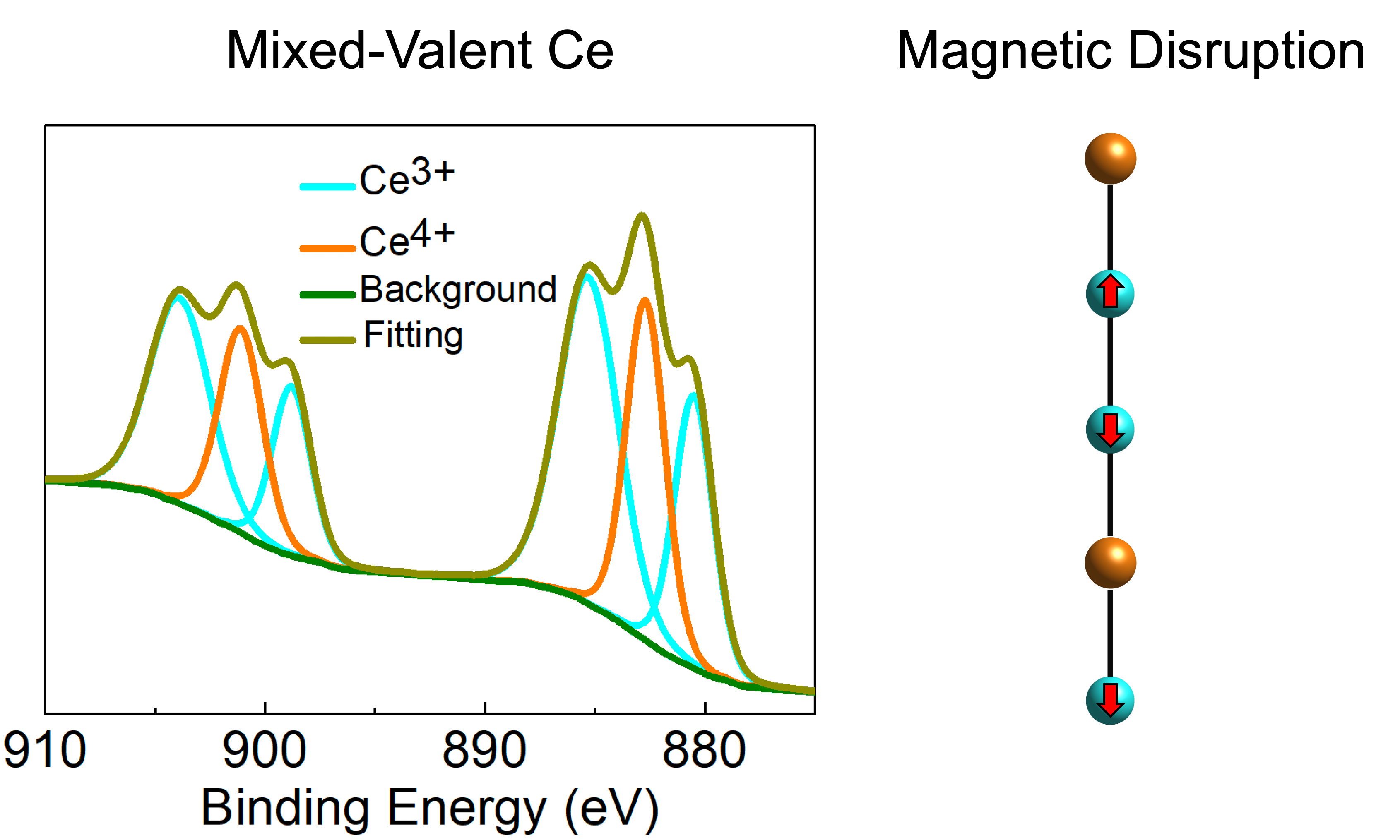}
\end{figure}

\end{document}


\maketitle
\tableofcontents
\newpage

\section{Energy Dispersive Spectroscopy (EDS) of CeFe$_2$Ga$_8$}

\begin{figure}[H]
    \centering
    \includegraphics[width=\textwidth]{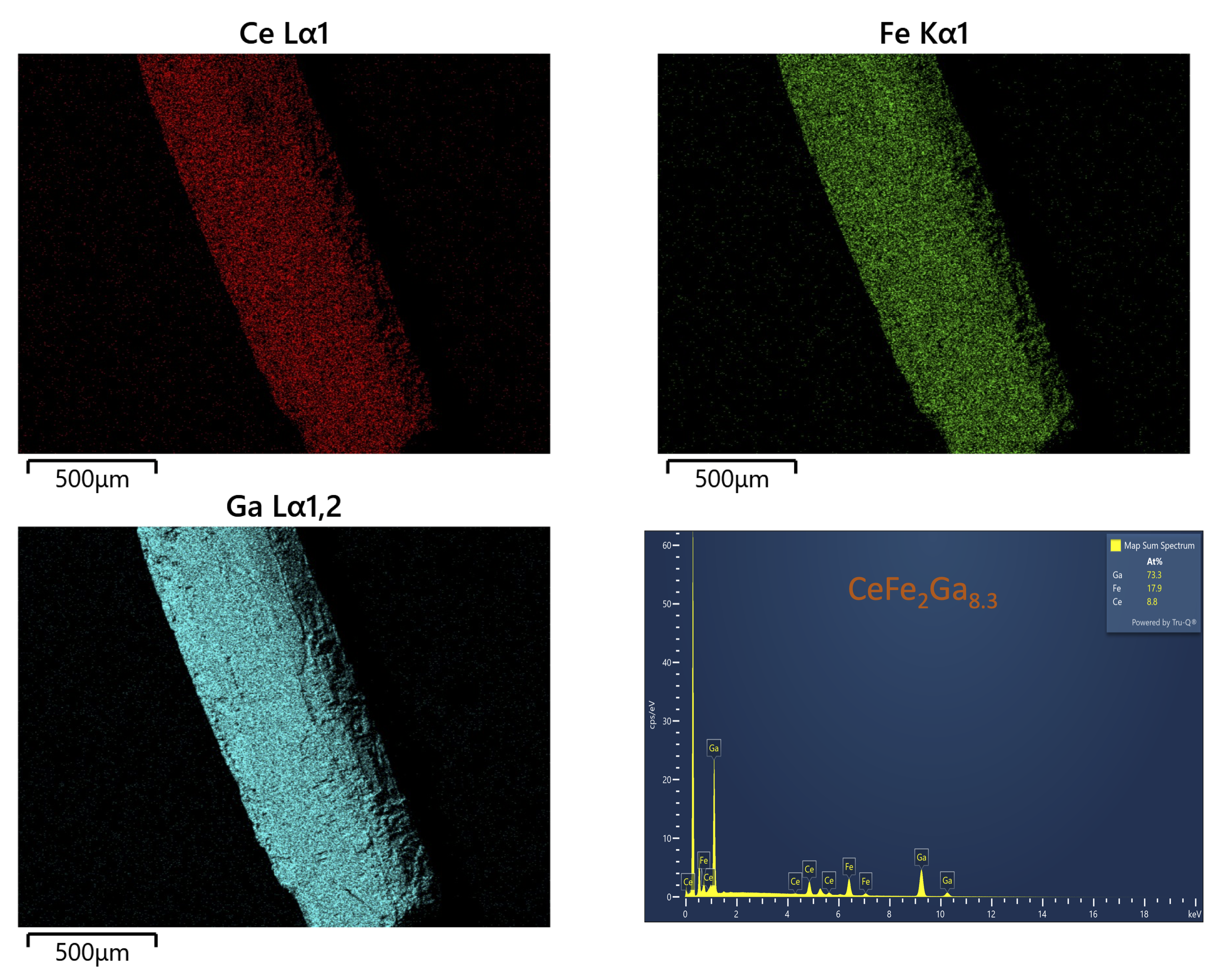}
    \captionsetup{labelformat=empty}
    \caption{\textbf{Figure S1.}EDS analysis of CeFe$_2$Ga$_8$ single crystal: elemental mapping showing homogeneous distribution of Ce (red), Fe (green), and Ga (cyan) across the crystal surface, with corresponding EDX spectrum showing characteristic peaks for all three elements. The compositional analysis yields CeFe$_{2}$Ga$_{8.3}$, confirming the expected stoichiometry.}
    \label{fig:edx}
\end{figure}

\section{ZFC-FC measurements of CeFe$_2$Ga$_8$ at 25 and 1000 Oe}

\begin{figure}[H]
    \centering
    \includegraphics[width=\textwidth]{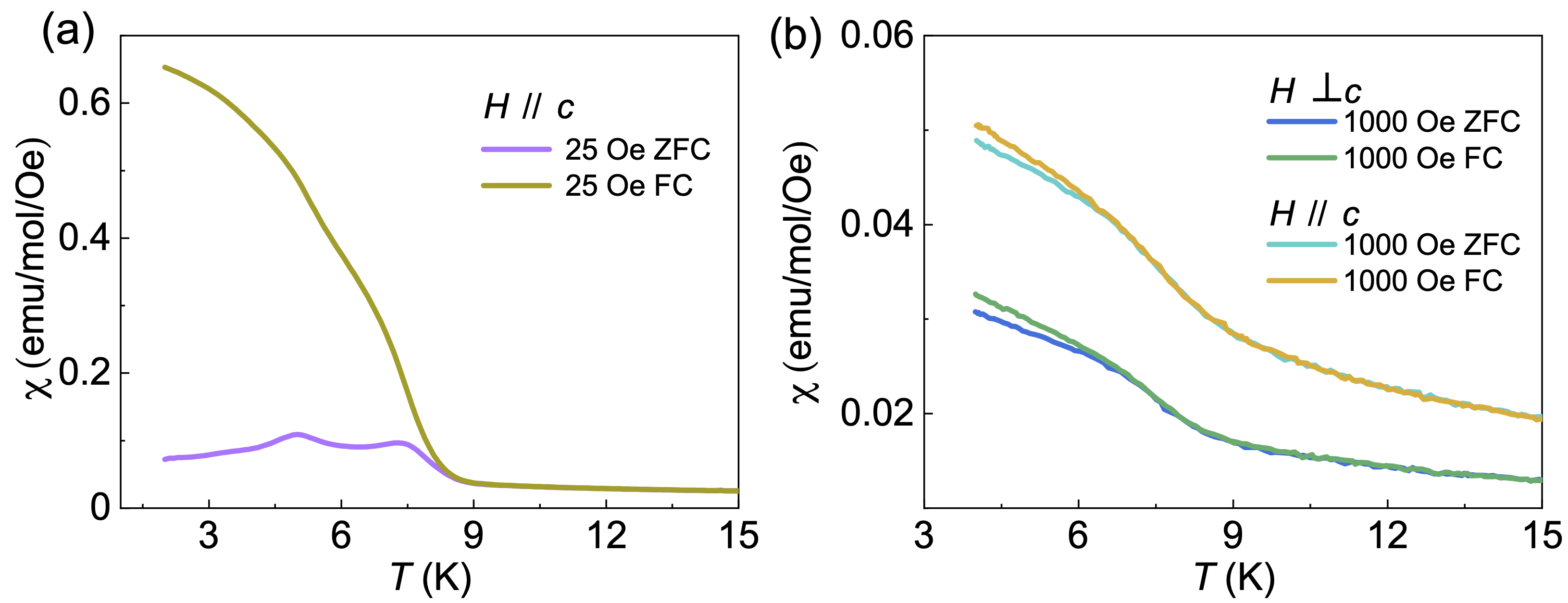}
    \captionsetup{labelformat=empty}
    \caption{\textbf{Figure S2.}Magnetic susceptibility measurements at 25 and 1000 Oe, highlighting the ZFC-FC divergence at both fields.}
    \label{fig:SIVSM}
\end{figure}


